\documentstyle[11pt,aaspp4,flushrt,epsfig,twoside]{article}
\newcommand{\ee}[2]{#1 \times 10^{#2}}%
\newcommand{\nuc}[1]{$\rm #1$}

\newcommand{\kB}{k_B}
\newcommand{\K}{\,{\rm K}}
\newcommand{\gram}{\,{\rm g}}
\newcommand{\secd}{\,{\rm s}}
\newcommand{\yr}{\,{\rm yr}}
\newcommand{\cm}{\,{\rm cm}}

\newcommand{\keV}{\,{\rm keV}}
\newcommand{\mueff}{\mu_{\rm eff}}
\newcommand{\fscr}{f_{\rm scr}}
\newcommand{\msun}{M_\odot}
\newcommand{\rsun}{R_\odot}
\newcommand{\Lsun}{L_\odot}
\newcommand{\Teff}{T_{\rm eff}}
\newcommand{\tdepl}{t_{\rm depl}}
\newcommand{\mrad}{0.5 \msun}
\newcommand{\Li}{[{\rm Li}]}
\begin{document}
\title{
Lithium Depletion in Fully Convective Pre-Main Sequence Stars
}
\author{
Lars Bildsten, Edward F. Brown, Christopher D. Matzner,  and
Greg Ushomirsky
} 
\affil{
Department of Physics and Department of Astronomy\\ 601
Campbell Hall \\ University of California, Berkeley, CA 94720
}
\authoremail{
bildsten@fire.berkeley.edu, ebrown@astron.berkeley.edu,
matzner@physics.berkeley.edu, gregus@fire.berkeley.edu
}
\begin{abstract}

We present an analytic calculation of the thermonuclear depletion of
lithium in contracting, fully convective, pre-main sequence stars of
mass $M\lesssim \mrad$. Previous numerical work relies on
still-uncertain physics (atmospheric opacities and convection, in
particular) to calculate the effective temperature as a unique
function of stellar mass. We assume that the star's effective
temperature, $\Teff$, is fixed during Hayashi contraction and allow
its actual value to be a free parameter constrained by observation.
Using this approximation, we compute lithium burning analytically and
explore the dependence of lithium depletion on $\Teff$, $M$, and
composition. Our calculations yield the radius, age, and luminosity of
a pre-main sequence star as a function of lithium depletion.  This
allows for more direct comparisons to observations of lithium depleted
stars.  Our results agree with those numerical calculations that
explicitly determine stellar structure during Hayashi contraction. In
agreement with Basri, Marcy, and Graham (1996), we show that the
absence of lithium in the Pleiades star HHJ 3 implies that it is older
than 100 Myr. We also suggest a generalized method for dating galactic
clusters younger than 100 Myr (i.e., those with contracting stars of
$M\gtrsim 0.08 M_\odot$) and for constraining the masses of lithium
depleted stars.
\end{abstract}


\begin{center}
\bf To appear in the Astrophysical Journal
\end{center}

\section{Introduction}

Lithium depletion in gravitationally contracting, fully convective
stars of mass $M<\msun$ (Hayashi \& Nakano 1963; Bodenheimer 1965) and
$M=\msun$ (Weymann \& Moore 1963; Ezer \& Cameron 1963) has been
studied for over thirty years. Motivated by lithium observations of
main sequence stars in young clusters and the halo (e.g., Soderblom
1995), many have calculated lithium depletion for stars of $M\gtrsim
\mrad$ both before and during the main sequence (Vauclair et al.\
1978; D'Antona \& Mazzitelli 1984; Proffitt \& Michaud 1989;
Vandenberg \& Poll 1989; Swenson, Stringfellow, \& Faulkner 1990;
Deliyannis, Demarque, \& Kawaler 1990). For stars with
$M\gtrsim\mrad$, most lithium burns after convection has halted in the
stellar core. Accurate predictions of lithium depletion then depend on
the temperature at the bottom of the retreating convective zone. The
location of the convective/radiative boundary and the amount of mixing
across it depend on opacity, treatment of convection, and rotation;
proper handling of these effects remains an open question.

Lower mass ($M\lesssim\mrad$) stars are fully convective during
lithium burning, which occurs before the star reaches the main
sequence.\footnote{The interstellar lithium abundance is so low
($N_{\rm Li}/N_{\rm H}\sim 10^{-9}$) that the energy released by its
fusion does not appreciably slow stellar contraction.  Throughout this
paper, we only consider \nuc{^7Li}, as \nuc{^6Li} always depletes
first and is less abundant in the local ISM (Lemoine, Ferlet, \&
Vidal-Madjar 1995).} The effective temperature $\Teff$ of a fully
convective star determines the contraction rate and is found by
matching the entropy at the interior to that at the photosphere (see
Stahler 1988 for a review of Hayashi contraction). For these low mass
stars with $\Teff\lesssim 4000 \ {\rm K}$, the opacities are still
uncertain, and there are still debates about the treatment of
convection.  These uncertainties have motivated many numerical
calculations of lithium depletion (Pozio 1991; Nelson, Rappaport, \&
Chiang 1993; D'Antona \& Mazzitelli 1994 (hereafter DM94); Chabrier,
Baraffe, \& Plez 1996). The differing input physics results in
different $\Teff$'s for the same stellar mass. For example, DM94 found
that, depending on the opacities used, $\Teff$ ranges from $3350\K$ to
$3640\K$ for a $0.2 \msun$ star.  Most calculations agree that $\Teff$
remains approximately constant during the fully convective contraction
phase.

In this paper, we present a different approach to calculating
pre-main sequence (or pre-brown dwarf) lithium depletion in
gravitationally contracting, low mass ($M\lesssim \mrad$) stars.
Rather than calculate $\Teff$, we calculate the dependence of lithium
depletion on $\Teff$. Given $\Teff$, $M$, and mean molecular weight
$\mu$, we can reproduce the results of prior works.  Moreover, our
approach allows the inferred  effective temperature to be used directly
in analyzing lithium depletion observations. Efficient convection
throughout the star allows it to be modeled as a fully mixed $n=3/2$
polytrope. Our analytic calculations then yield the age, radius, and
luminosity at a given level of lithium depletion. Our results
(equations [\ref{eq:fit-allmass}] and [\ref{eq:fit-highmass}]) can be
used to constrain the mass and age of a star from its lithium
abundance. We apply these same techniques to depletion of beryllium
and boron in Ushomirsky et al.\ (1997). 

\section{Contraction and Lithium Burning in Fully Convective Pre-Main Sequence
 Stars }
 
A number of authors have undertaken detailed numerical evolutionary
calculations of contracting pre-main sequence stars (see Burrows \&
Liebert 1993 for a review).  Low mass stars ($M\lesssim\mrad$)
remain completely convective at least through the end of lithium
burning (DM94). For an ideal gas ($P=\rho N_A \kB T/\mu$ where $N_A$
is Avogadro's number) the adiabatic relation is $P\propto\rho^{5/3}$,
so that an $n=3/2$ polytrope describes the stellar structure.  During
later stages of contraction, electron degeneracy modifies the equation
of state by reducing the central temperature below its non-degenerate
value. However, partial degeneracy does not alter the polytropic
structure.  The central density and temperature are
\begin{equation}\label{eq:polytropes} 
\label{rho_c-polytrope}
\rho_c = 8.44\left(\frac{M}{\msun}\right)
\left(\frac{\rsun}{R}\right)^{3} \gram\cm^{-3},\qquad
T_{c} = 7.41\times 10^6 
\left(\frac{\mueff}{0.6}\right)
\left(\frac{M}{\msun}\right)
\left(\frac{\rsun}{R}\right) \K,
\end{equation}
where 
\begin{equation}
\mueff \equiv \frac{\rho N_A \kB T}{P} \le \mu
\end{equation}
accounts for the deviation in electron pressure from that of an ideal
gas.  We neglect Coulomb and ionization corrections to the equation of
state.  Except during deuterium burning, gravitational contraction
powers the star's luminosity, $L = 4\pi R^2 \sigma_{\rm SB}\Teff{}^4 =
-(3 GM^2/7R^2)(dR/dt)$, which is independent of the degree of degeneracy.

We define the time coordinate $t$ under the assumption that $\Teff$
is constant during contraction from a formally infinite radius. This
time therefore differs from chronological age because of the deuterium
burning phase and the initial radius on the theoretical stellar
birthline (\cite{sta88}). Lithium depletion, however, occurs long
(10--100 Myr) after these events, so that if the effective temperature
used is correct at the time of lithium depletion, $t$ differs only 
slightly from chronological age. Gravitational contraction then gives
the stellar radius and luminosity as functions of time as
\begin{eqnarray}\label{eq:rstar}
\frac{R}{\rsun} &=& 0.850\left(\frac{M}{0.1 \msun}\right)^{2/3}
	\left(\frac{3000 \K}{\Teff}\right)^{4/3}
\left(\frac{{\rm Myr}}{t}\right)^{1/3}, \\
\label{eq:lstar}
\frac{L}{\Lsun} &=& \ee{5.25}{-2}\left(\frac{M}{0.1 \msun}\right)^{4/3}
	\left(\frac{\Teff}{3000\K}\right)^{4/3}
\left(\frac{{\rm Myr}}{t}\right)^{2/3}. 
\end{eqnarray}
We define the contraction timescale in terms of the central
temperature,
\begin{equation} \label{eq:tcont}
t_{\rm cont}\equiv -\frac{R}{dR/dt} = 115 \left(\frac{3000 \K}{\Teff}\right)^4
\left(\frac{0.1 \msun}{M}\right)
\left(\frac{0.6}{\mueff}\right)^3
\left(\frac{T_c}{3\times 10^6\K}\right)^3{\rm Myr}= 3t,
\end{equation}
which we then compare to the timescale for lithium destruction due to the
reaction \nuc{^7Li(p,\alpha)\,^4He}.  Over the range of temperatures
($T_6\equiv (T/10^6\K) <6$) appropriate for this work, the reaction
rate is $N_A\langle\sigma v\rangle=S \fscr
T_6^{-2/3}\exp(-aT_6^{-1/3})
\cm^3\secd^{-1}\gram^{-1}$ where $S=6.4\times 10^{10}$ and $a=84.72$
(Caughlan \& Fowler 1988) and $\fscr$ is the screening correction
factor (\cite{sal69}). Raimann (1993) recently discussed new
experimental results at low energies ($\approx 11$--$13\keV$) and
updated $S$ to $7.2\times 10^{10}$.

Lithium is depleted when the local nuclear destruction time, $t_{\rm
dest}\equiv m_p/X\rho\langle \sigma v\rangle$ ($X$ is the hydrogen
mass fraction and $m_p$ is the proton mass), becomes comparable to
$t_{\rm cont}$. Although the full calculation involves integrating
over the star (\S 3), one obtains a flavor of the calculation by
evaluating $t_{\rm dest}$ at the center of the star (with $\fscr=1$
and $\mueff=\mu$) and equating it with $t_{\rm cont}$.  This gives a
relation for the central temperature $T_{c6}\equiv T_c/10^6 \K$ at the
time of depletion,
\begin{equation}\label{eq:trans}
\frac{a}{T_{c6}{}^{1/3}} = 32.9 + \ln(S) -
3\ln \left(\frac{M}{0.1 \msun}\right)-
4 \ln \left(\frac{\Teff}{3000 \K} \right) -
6\ln\left(\frac{\mu}{0.6}\right)
+ \frac{16}{3} \ln T_{c6}.
\end{equation}
For example, if $\Teff=3500 \K$ and $M=0.5\msun$, then \nuc{^7Li}
depletes when $T_{c6}=3.04$. The temperature is only slightly
different for other masses. For these temperatures, the reaction rate
is extremely temperature sensitive ($\propto T^{20}$), so that $t_{\rm
dest}$ decreases rapidly as the temperature increases.  Hence the
transition from no depletion to full depletion occurs rapidly as long
as the central temperature reaches these values.  This does not occur
for very low masses ($\lesssim 0.06 \msun$), because electron
degeneracy pressure dominates the equation of state (Pozio 1991;
Magazz\`u, Mart\'{\i}n, \& Rebolo 1993; Nelson et al.\ 1993).  Also,
stars with $M \gtrsim \mrad $ develop radiative cores prior to
depletion.  We only discuss stars safely between these limits.  Since
this depletion occurs on a contraction timescale (10--100 Myr), the
central temperature during depletion is higher ($\approx (3$--$4)
\times 10^6 \K$) than the temperature needed ($2.4\times 10^6 \ {\rm
K}$) at the base of a main sequence star's convective zone to deplete
\nuc{^7Li} while the star is on the main sequence.

The narrow range of burning temperatures allows the solution of
equation (\ref{eq:trans}) to be approximated as a power law. For the
non-degenerate case, $T_c$ is $\propto M^{1/8} \Teff{}^{1/6}\mu^{1/4}$
at the time of depletion, and $\tdepl$ is $\propto T_{\rm
eff}{}^{-7/2}M^{-5/8}$.  The radius at depletion, $R \propto M^{7/8}
\mu^{3/4} \Teff{}^{-1/6}$, is relatively insensitive to $\Teff$ but
nearly proportional to mass. These scalings, which are modified only
slightly in the full calculation (\S 3), provide an intuitive picture
of lithium burning and a means to evaluate observations (\S 4) once
the prefactors are known.

\section{Full Calculation of Lithium Depletion}

Since lithium burns near the center of the star, depleting lithium
throughout the star requires mixing lithium-poor fluid outward and
lithium-rich fluid inward. For efficient convection, the mixing
timescale ($\sim 10$--$100\yr)$ is much shorter than both $t_{\rm
cont}$ and $t_{\rm dest}$; convective mixing keeps the lithium to
hydrogen ratio, $f$, fixed throughout the star as the total lithium
content decreases. Thus we write the global depletion rate as 
\begin{equation}
M {df\over dt}= -{Xf\over m_p} 
\int_0^M \rho\langle\sigma v\rangle dM.
\end{equation} 
Changing to spatial variables and using the thermonuclear rate defined
earlier, we obtain
\begin{equation}
\frac{d}{dt} \ln f = -\frac{4 \pi X}{N_A m_p M}
\int_0^R r^2 \rho^2 S\fscr
T_6{}^{-2/3}\exp\left(-\frac{a}{T_6{}^{1/3}}\right) \, dr. 
\end{equation}
The temperature sensitivity of the nuclear reaction allows us to
expand $T$ and $\rho$ about their central values. 
Integrating the lowest order terms (see Ushomirsky et al.\ 1997 for a
discussion of the small errors introduced by this approximation) 
yields 
\begin{equation}\label{eq:dt-depletion} 
\frac{d}{dt} \ln f 
= - 18.0
\left(\frac{X}{0.70}\right)
\left(\frac{0.6}{\mueff}\right)^{3}
\left(\frac{0.1\msun}{M}\right)^{2}
S\fscr a^7 \alpha^{-17/2} 
\left(1-\frac{21}{2\alpha}\right) e^{-\alpha},
\end{equation}
where $\alpha \equiv a T_{c6}{}^{-1/3}$ is a convenient representation
of the central temperature. Using equation (\ref{eq:tcont}) and the
fact that $d\alpha/dR=\alpha/3R$, we characterize the stellar state by
$\alpha$ and integrate from $\alpha= \infty$ and initial abundance
$f_\circ$ to find the depletion $W$ as a function of $\alpha$,
\begin{equation}\label{eq:exact-depletion}
W \equiv \ln \left({f_\circ\over f}\right)= \ee{7.70}{15}S\fscr a^{16}
\left(\frac{X}{0.70} \right)
\left(\frac{0.6}{\mueff}\right)^{6}
\left(\frac{3000 \K}{\Teff}\right)^4
\left(\frac{0.1 \msun}{M}\right)^{3} g(\alpha),
\end{equation}
where $g(\alpha)=\alpha^{-37/2}e^{-\alpha}-29
\Gamma\left(-37/2,\alpha\right)$ and $\Gamma(-37/2, \alpha)$ is an
incomplete gamma function. This transcendental equation is similar to
that found from the simple arguments in \S 2. For a given depletion
$W$, the solution $\alpha(W)$ determines $T_c$. The radius,
luminosity, and age then follow from equations
(\ref{eq:rstar})--(\ref{eq:tcont}). Our treatment of degeneracy and
screening in equation (\ref{eq:exact-depletion}) is approximate in the
sense that we neglect the slight variation of $\mueff$ and $\fscr$
with $\alpha$.  Instead, we evaluate their values at the time of
depletion to solve the resulting transcendental equation. We discuss
this process in more detail in a forthcoming paper (Ushomirsky et al.\
1997).

Although solving the depletion equation (\ref{eq:exact-depletion})
numerically is straightforward, an analytic fitting formula for the
time (and therefore the radius and luminosity from equations
[\ref{eq:rstar}] and [\ref{eq:lstar}]) as a function of depletion is
also helpful.  The age at a given depletion (we use Raimann's [1993]
rates; using Caughlan's and Fowler's [1988] rates changes $\tdepl$ by
no more than a few percent) for $2000 \K < \Teff < 4000 \K$, $0.075
M_\odot < M < 0.5 M_\odot$, and $0.65 < X < 0.75$ is
\begin{eqnarray}\label{eq:fit-allmass}
\tdepl &=&54.1\left(\frac{0.1 \msun}{M}\right)^{0.715}
	\left(\frac{3000\K}{\Teff}\right)^{3.51}
	\left(\frac{0.6}{\mu}\right)^{1.98}
	\left(\frac{W}{\ln 2}\right)^{0.121} \\ \nonumber
  &&	\times
	\left[
		1+0.117\left(\frac{0.1 \msun}{M}\right)^{6.39}
		\left(\frac{\Teff}{3000\K}\right)^{0.828}
		\left(\frac{0.6}{\mu}\right)^{11.8}
		\left(\frac{W}{\ln 2}\right)^{0.204}
	\right] {\rm Myr}.
\end{eqnarray}
The first line of equation (\ref{eq:fit-allmass}) follows from the
simple arguments of \S 2, while the expression in square brackets
accounts for the onset of degeneracy in lower-mass stars. A simpler
fit,
\begin{equation}\label{eq:fit-highmass}
\tdepl = 50.7\left(\frac{0.1 \msun}{M}\right)^{0.663}
	\left(\frac{3000 \K}{\Teff}\right)^{3.50}
	\left(\frac{0.6}{\mu}\right)^{2.09} \left(\frac{W}{\ln
	2}\right)^{0.124} {\rm Myr}, 
\end{equation}
is obtained if degeneracy is not important during depletion
($0.2\msun\lesssim M \lesssim 0.5\msun$). These fits agree with
Chabrier et al.\ (1996, Table 1) for masses $0.08\msun\le M\le
0.5\msun$ at 50\% (99\%) depletion to better than 4\% (8\%), 7\%
(17\%), and 12\% (25\%) for the radius, luminosity, and age,
respectively.\footnote{Percentages reported here are {\it maximum}
deviations of our results (equations [\ref{eq:fit-allmass}] and
[\ref{eq:fit-highmass}]) from the stated references.}  We also
reproduce the results in Tables 5--8 of DM94 for $0.07\msun\le M\le
0.3\msun$ to better than 3\%, 7\%, and 15\% for the radius,
luminosity, and age, respectively.
 
The strong temperature dependence of the burning rate makes $T_c$ at
depletion very insensitive to our approximations. The central
temperature is thus the most accurately determined quantity. Indeed,
our $T_c$'s at a given depletion level deviate from those reported in
Tables 5--8 of DM94 by less than 2\%\ (Chabrier et al.\ [1996] did not
report $T_c$). Our neglect of Coulomb and ionization effects
(\cite{berkel}) introduces a small error in mapping $T_c$ to $R$
(roughly, $T_c\propto M/R$). These errors then propagate into the age
and luminosity determination at a given depletion level. Since the age
is $\propto T_c^3$ (see equation [\ref{eq:tcont}]), the relative
deviations in $t$ are roughly three times larger than those in
$R$. Moreover, any changes in $\Teff$ during contraction cause
additional uncertainty (typically up to $5\%$, see Ushomirsky et
al.\ 1997) in calculating how long a star takes to contract to a
certain radius.

In summary, the absolute error of age determination in our calculation
is no more than $20$--$25\%$ under any conditions, and the average
discrepancy with published results is about half that amount.  In
light of these uncertainties, it is important to note that, because
the age at a level of depletion is $t\propto 1/\Teff^{3.5}$, the
typical $5\%$ observational errors in $\Teff$ lead to uncertainties in
age that are comparable to the deviations of our results from the
detailed numerical ones. Our results therefore are adequate for
observational work, and have the added advantage of enabling the
observer to use the derived $\Teff$ and $L$ in conjunction with
lithium observations to constrain the stellar mass.  This is in
contrast to using detailed evolutionary tracks to infer the mass from
the star's location on the HR diagram. As we show in the following
section, the inferred mass can be in conflict with lithium
depletion. In addition (insofar as $\Teff$ is nearly constant during
contraction), our results can serve as benchmarks in comparing and
evaluating detailed numerical calculations.
\begin{figure}[hbp]
\centering{\epsfig{file=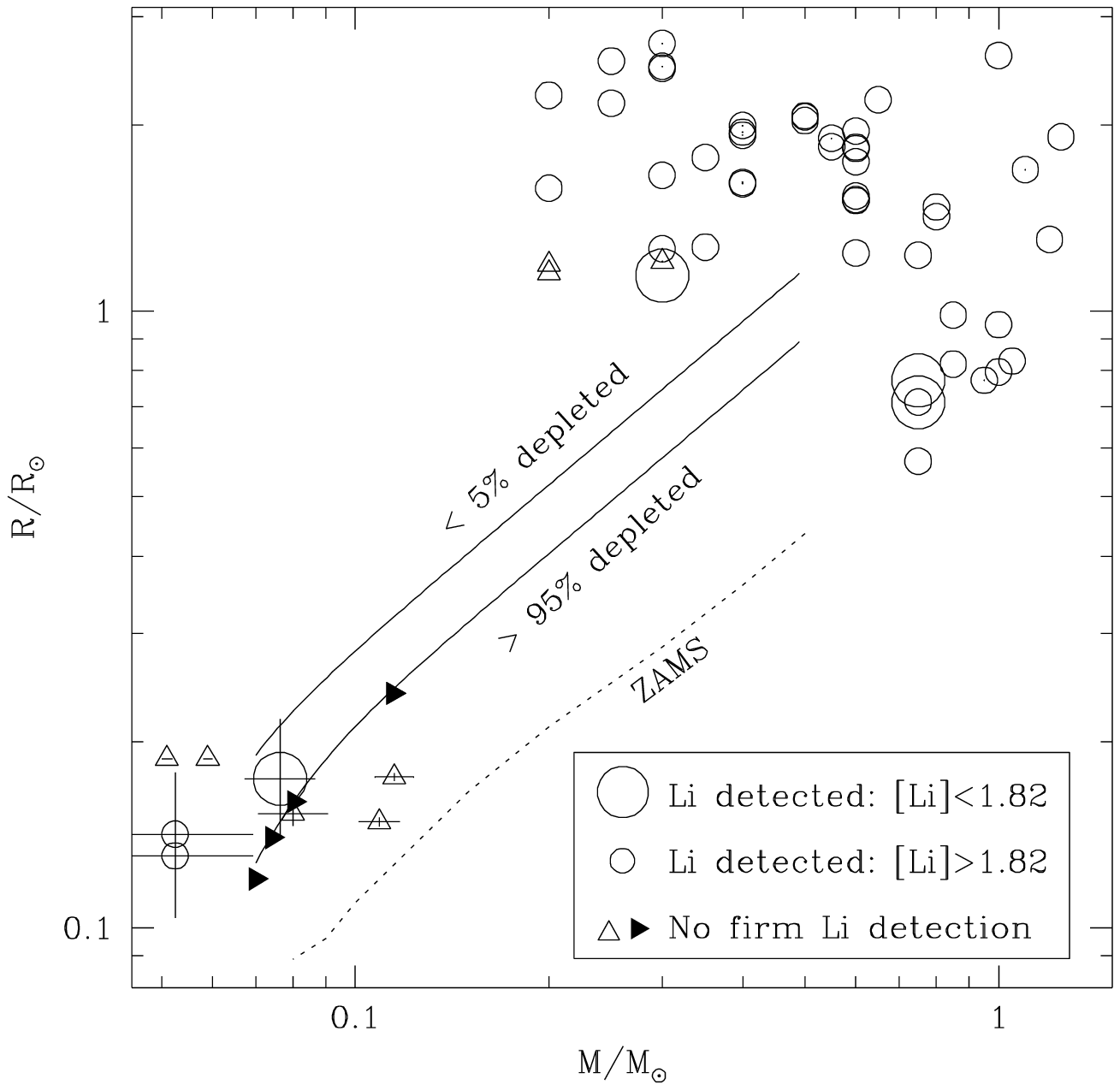,width=\textwidth}}
\large
\renewcommand{\baselinestretch}{0.7}
\footnotesize
\vspace*{-20pt}
\caption{\protect{\footnotesize
A compilation of published values (\protect\cite{bur89};
\protect\cite{mag91}; \protect\cite{mag93}; \protect\cite{mar94a};
\protect\cite{reb96}) of $M$, $R$, and $N_{\rm Li}$ for pre-main
sequence stars. Small circles represent stars reported to have
$\Li\equiv 12+\log(N_{\rm Li}/N_{\rm H}) >1.82$, which corresponds to
an abundance of more than $5\%$ of the Population I value ($\Li =
3.1$). Large circles correspond to stars with $\Li < 1.82$.
Triangles, both filled and empty, indicate stars with no firm lithium
detection: either no lithium line is detected, or only an upper bound
is reported for $N_{\rm Li}$.  Where applicable, we have indicated
reported uncertainties in mass and radius.  The two solid lines are
contours of constant depletion calculated from equation 
(\ref{eq:fit-allmass}) with $\mu=0.6$ and $X=0.7$ and with an
effective temperature law $\Teff = 4000(M/\msun)^{1/7}\K$; the
position and shape of the depletion contours are insensitive to these
choices. The solid lines end at $M=\mrad$; above this mass stellar cores
are not convective at the depletion time (\protect\cite{dan94}). At
lower left are a series of points (filled right-pointing triangles)
for which no lithium was detected and no mass was reported (cf.\ Table
\ref{t:observations}).  Taking the non-detection of lithium to mean
that these stars are more than 95\% depleted, we solve for their
minimum masses. The dotted line is the theoretical zero-age main
sequence for $0.08\msun\le M\le 0.5\msun$ (\protect\cite{dan94}).
\label{fig:obs}}}
\end{figure}

\section{
Comparison to Observations of Lithium Depleted Pre-Main Sequence Stars}

The radii of pre-main sequence stars are typically inferred from
$\Teff$ and $L$, but (except in binaries) masses must be estimated by
relating colors to a grid of computed evolutionary tracks.  There are
potentially significant uncertainties, both observational and
theoretical, associated with inferring masses and radii of stars from
these observations. On the other hand, the presence or absence of
lithium is an indirect indicator of the central temperature of the
star. Since lithium does not deplete until a characteristic central
temperature ($T_c\sim\ee{3$--$4}{6}\K$) is reached, the lithium
abundance can be used to constrain the mass to radius ratio (cf.\
equations [\ref{eq:rstar}] and [\ref{eq:fit-allmass}]). Age
determination is, in some sense, secondary to measuring the radius.

Since the central temperature is $\propto M/R$, the additional
constraint provided by the lithium abundance measurements (that are
sensitive to the derived $\Teff$) is most easily exploited by plotting
stellar radii versus masses. Figure
\ref{fig:obs} is a compilation of published mass and radius values of
pre-main sequence stars where lithium abundances have been estimated
or constrained. We divide the stars into high (small circles), low
(large circles), and undetectable (triangle) lithium content bins (see
caption).  Also plotted are lines of constant depletion, which
correspond to nearly constant central temperature, i.e., $R\propto
M$. These lines are relatively insensitive to the effective
temperature and divide the graph into an undepleted, a depleting, and
a depleted region. The narrowness of the depleting region results from
the reaction's strong temperature sensitivity. We do not continue our
lines to high masses where the stars develop radiative cores prior to
depletion, nor to low masses where the stars become strongly
degenerate.

A few stars appear depleted but lie above the depletion lines. Since
central burning cannot have caused such a low lithium abundance,
either the star was born with an anomalously low abundance or the
reported masses and/or radii are in error.  A challenge to our
assumption of complete mixing would be a detection of lithium in a
star lying beneath the depletion lines. We have found no such
measurement for $M<\mrad$.  We also display in Figure \ref{fig:obs}
those stars for which no lithium is detected and no mass is reported
(filled triangles).  We interpret the lack of lithium to imply that
more than 95\%\ has burned and use the inferred radii to constrain
their masses (see Table 1). Determinations of $\Teff$ from the
observed colors are still highly uncertain and depend on the
temperature scale and model atmosphere used. We have not attempted to
convert observed colors to a single temperature scale, but have just
used the quoted results.  However, our figure depends only on the
value of $\Teff$ and not the physics needed to infer it.
\begin{deluxetable}{lrrrrrl}
\tablecolumns{7}
\tablewidth{352pt}
\tablecaption{%
Radius and Mass of Lithium Depleted Pre-Main Sequence Stars
\label{t:observations}}
\tablehead{
\colhead{} & \multicolumn{2}{c}{Observed} & \colhead{} &
\multicolumn{2}{c}{Inferred} & \colhead{}\\
\cline{2-3}\cline{5-6}\\
\colhead{Star} & \colhead{$T_{\rm eff}({\rm K})$} & \colhead{$\log(L/L_\odot)$}
&\colhead{} & \colhead{$R/R_\odot$} & \colhead{$M_{\rm min}/M_\odot$} &
\colhead{Ref.} }
\startdata
HHJ 10 & 3120 & -2.8 & & 0.14\phn & 0.074 & 1,2,(6) \nl
HHJ 36 & 2825 & -2.5 & & 0.24\phn & 0.11\phn & 3,2,(6) \nl
LHS 248 & 2775 & -2.5 & & 0.24\phn & 0.11\phn & 4,(1) \nl
GL 406  & 2600 & -3.0 & & 0.16\phn & 0.078 & 5,(7) \nl
GL 569B & 2773 & -3.1 & & 0.12\phn & 0.069 & 5,(7) \nl
\enddata
\tablecomments{\protect{\footnotesize
The 5\%\ uncertainties in $T_{\rm eff}$ correspond to a 10\%
uncertainties in radii and uncertainties of no greater than 10\% in
the minimum masses. The lithium non-detection is reported in the
reference in parentheses. For stars with two references, the first is
for effective temperature; the second, for luminosity. It is important
to note that we have not calibrated the effective temperatures to a
common scale. To quote our anonymous referee, ``the pedigree of any
temperature derived from observations cannot be ignored.'' The
systematic errors between different $\Teff$ scales can be 
as much as 10\%. }}
\tablerefs{\protect{\footnotesize
(1) Mart\'{\i}n et al.\ 1994b; (2) Stauffer et al.\ 1995;
(3) Steele et al.\ 1995; (4) Bessel \& Stringfellow 1993; (5) Burrows,
Hubbard, \& Lunine 1989; (6) Oppenheimer et al.\ 1996; (7) Magazz\`u
et al.\ 1993}}
\end{deluxetable}

Figure 1 also clearly shows that there is a dearth of observations of pre-main
sequence stars near the depleting region with masses
$0.1M_\odot<M<\mrad$. {\em Observations in this mass range are important, as
they would test the mixing assumption and may yield, independently of
atmospheric physics, $\Teff$ as a function of mass.}

As mentioned above, there are significant uncertainties in relating
the observed colors to effective temperatures for low-mass stars, and
in using the inferred $\Teff$'s and evolutionary models to determine ages
of stars. Nevertheless, we can still bound the stellar age and mass
without relying on exact knowledge of $\Teff$, as long as $L$ is well
determined and lithium is unobserved\footnote{Similarly, a detection of 
lithium coupled with a lower bound on $\Teff$ yields the maximum age.}
down to some detection limit $W_\circ$. For a trial value for the 
age of the star, $t$, the contraction equation (\ref{eq:lstar}) yields
$M \propto t^{1/2}/\Teff$.  Substituting this relation into the
depletion formula, equations (\ref{eq:fit-allmass}) or
(\ref{eq:fit-highmass}), we obtain the time $\tdepl$ at which this
star is depleted to $W_\circ$. For consistency with the lack of lithium,
we demand $t \ge \tdepl$, with equality yielding the lower limit
on age for a given $\Teff$. For high-mass stars, $t$ is $\propto
\Teff^{-2.1}$, and the upper bound on $\Teff$ determines the minimum
age. For lower mass stars, $t(\Teff)$ always has a minimum (due to the
effects of degeneracy),
\begin{equation}\label{eq:tmin}
t_{\rm min}= 58.3\left(\frac{10^{-2.5}\Lsun}{L}\right)^{0.922}
        \left(\frac{0.6}{\mu}\right)^{2.52} \left(\frac{W_\circ}{\ln
        2}\right)^{0.0769} \,{\rm Myr},
\end{equation}
which is the lower bound on age independent of $\Teff$  for $-3.50 <
\log(L/L_\odot) < -2.32$. Because 
the minimum in $\tdepl$ is due to the effects of degeneracy, this
technique is independent of $\Teff$ only for low-mass stars; this is
fortunate because for these stars $\Teff$ is more uncertain.

 We apply equation (\ref{eq:tmin}) to the lithium-depleted star HHJ 3
in the Pleiades ($\log[L/\Lsun]=-2.78\pm 0.05$ [\cite{bas96}]). If the
absence of \nuc{^7Li} in its spectrum implies that this star is more
than 99\%\ depleted, then our calculations imply that it must be older
than 100 Myr for normal composition ($X=0.7$), regardless of its
effective temperature or mass (i.e., regardless of an evolutionary
track.) Basri et al.\ (1996) reached the same conclusion by using the
evolutionary tracks of Nelson et al.\ (1993). That our estimates agree
is a consequence of the age constraint (in this low-mass range)
depending mostly on $L$ and being nearly independent of $\Teff$.  As
noted by Basri et al.\ (1996), this age agrees with the age of the
Pleiades inferred from main sequence turn-off {\it with\/} convective
overshoot (\cite{mazzei}; \cite{maeder}). In contrast, to share the
60--70 Myr age of the Pleiades inferred from main sequence turn-off
{\it without\/} convective overshoot (\cite{mazzei}), HHJ 3 must have
either a hydrogen content $X\sim 0.5$ (very unlikely) or a luminosity
50\% higher than that reported.
\begin{figure}[hbp]
\centering{\epsfig{file=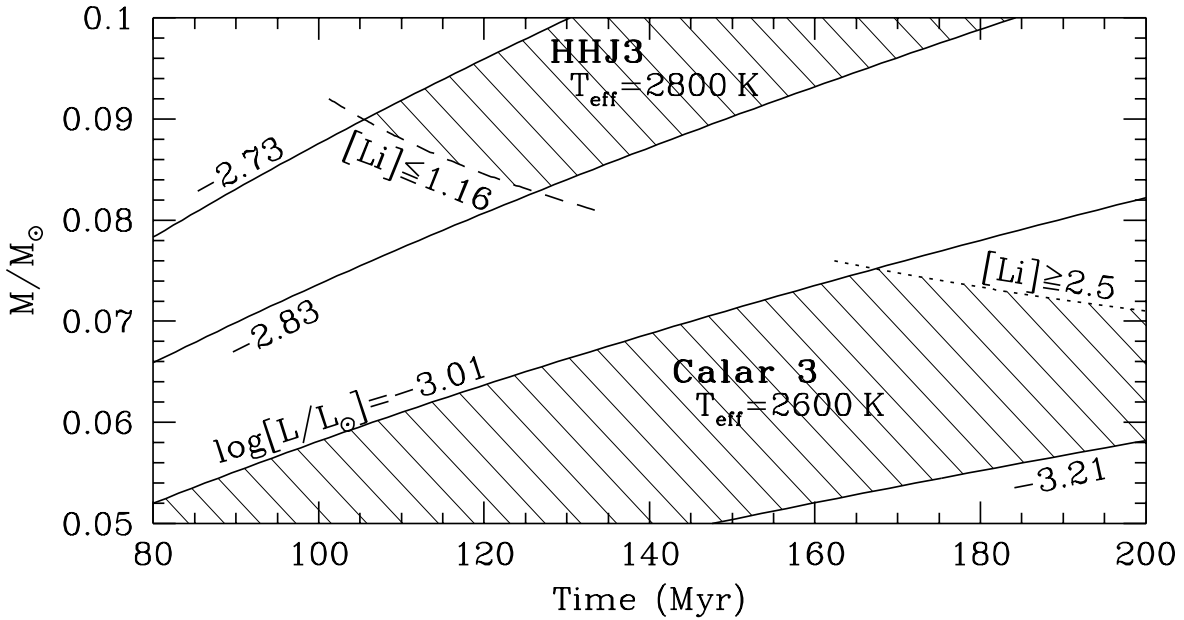,width=\hsize}}
\large
\renewcommand{\baselinestretch}{0.7}
\footnotesize
\vspace*{-20pt}
\caption{\protect{\footnotesize
An example of using the formulae for depletion time (eq.\
[\protect\ref{eq:fit-allmass}]), radius (eq.\
[\protect\ref{eq:rstar}]), and luminosity (eq.\
[\protect\ref{eq:lstar}]) to bound the age of the Pleiades.  For both
HHJ 3 and Calar 3, we plot a contour of constant lithium abundance
(dashed for HHJ 3; dotted for Calar 3) and two constant luminosity
contours (solid lines).  These constraints bound for each star a
region of $M$--$t$ space (shaded areas).  HHJ 3 has an effective
temperature $\Teff = 2800\K$ (\protect\cite{ste95}), a luminosity
$-2.83<\log(L/\Lsun) < -2.73$ (\protect\cite{bas96}), and $\Li <
1.16$.  For Calar 3, the region of $M$--$t$ space is bounded by $\Teff
= 2600\K$, $-3.21<\log(L/\Lsun)<-3.01$, and $\Li > 2.5$.  The overlap
in $t$ of the shaded areas sets a minimum age (105 Myr) and maximum
age (not shown) of the Pleiades.  Within this age range, each star
then has a minimum and maximum allowed mass.  (\protect\cite{reb96}).
\label{fig:pleiades}}}
\end{figure}

If both $L$ and $\Teff$ are known, the absence (presence) of lithium
sets a meaningful lower (upper) limit on both the mass and the age of
an individual star.  This method had its first practical application
in the work of Basri et al.\ (1996), who used it to find a new age for
the Pleiades and argued that the Pleiades' faintest members are
substellar objects. Instead of plotting isochrones and evolutionary
tracks on an $L$--$\Teff$ diagram, we find it more convenient to
directly relate the two unknown quantities, mass and age, on an
$M$--$t$ plot (Figure 2). On this plot, the observed luminosity is
represented by a line with positive slope found from the contraction
relation (\ref{eq:lstar}), while the limit on lithium abundance is
represented by a line with negative slope inferred from the depletion
formula (\ref{eq:fit-allmass}). For a given $\Teff$, the star must lie
in a swath of the $M$--$t$ plane defined by the estimated luminosity
range. The lithium abundance constraint intersects this swath, setting
limits on age and mass. 

As an example, the hatching in Figure \ref{fig:pleiades} denotes the
allowed regions of the Pleiads HHJ 3 and Calar 3 (which shows lithium
and appears undepleted) (Basri, Marcy, \& Graham 1996; Rebolo et al.\
1996). Assuming that the stars are coeval, Calar 3 must have a mass
$<0.075 \msun$, and HHJ 3 must have a mass $>0.08 \msun$, if the
effective temperatures (\cite{ste95}) are correctly measured. The
recent detection of lithium in the Pleiades star PPl~15 ($\log[N_{\rm
Li}/N_{\rm H}]\ge -10.84$ [\cite{bas96}]), which has
$\Teff=2800\pm150\K$ and $\log[L/\Lsun]=-2.80\pm 0.10$ (\cite{reb96}),
provides a better upper bound on the age of the Pleiades than Calar
3. In particular, we find that the maximum age of PPl~15 (and hence of
the Pleiades) is $~145$~Myr while its mass is constrained to lie in
the range $0.07\msun$--$0.09\msun$. 

\section{Conclusions}

Assuming that a contracting star is fully mixed and that the time it
spends prior to Hayashi contraction is negligible compared to typical
depletion times, we have derived simple analytical relations
(equations [\ref{eq:rstar}], [\ref{eq:lstar}], [\ref{eq:fit-allmass}],
and [\ref{eq:fit-highmass}]) for the radius, luminosity, and age of a
star as a function of lithium depletion. These formulae demonstrate
the dependence of the time of lithium depletion on the mass,
composition, and effective temperature of the star. Reasonable
agreement with other published theoretical calculations supports the
use of our results to evaluate observations.

We outline a method for using the observed $\Teff$ and $L$ of both
depleted and undepleted stars in a cluster to constrain its age.  This
method (complementary to that used by Basri et al.\ [1996] to date the
Pleiades) is also applicable to clusters of age 10--100 Myr, where
higher mass stars (up to $\mrad$), are presently depleting
lithium. These stars are relatively luminous ($L\gtrsim 10^{-2}
L_\odot$) when depleting lithium and are hence more easily observed.
For example, a $0.3 M_\odot$ star will have depleted 50\% of its
lithium at about 16 Myr, when it has a luminosity $0.04 L_\odot$ (for
$\Teff=3300\K$). Because stars with a range of masses will
deplete simultaneously (the mass of a 99\%\ depleted star is
approximately 1.4 times the mass of a 50\%\ depleted star),
observations of stars within this mass range will constrain the
cluster's age and the relative ordering of stellar masses.
In a young cluster, the most
stringent bounds are obtained by observing the dimmest depleted star
and the brightest undepleted star.

We have assumed that a fully convective star mixes material from the
core to the photosphere faster than it contracts. Until a fully
convective pre-main sequence star is observed that has both lithium
and a central temperature hotter than the maximum burning temperature
($4.4\times 10^6 \K$ for a $\mrad$ star), this approximation goes
unchallenged. In this sense, we have calculated the earliest time for
depletion in the contracting star. Incomplete mixing can only delay
the time of depletion to a given level.

  We thank foremost the graduate students in UC Berkeley's IDS 252
(Spring 1996) class who initially worked on this problem, especially
Vadim Hilliard and Peter Schroeder. We also thank Isabelle Baraffe,
Gibor Basri, Gilles Chabrier, Lynne Hillenbrand, Geoff Marcy, Chris
McKee, and John Monnier for helpful discussions. We thank the anonymous
referees for constructive comments.  L. B. acknowledges
support as an Alfred P. Sloan Foundation fellow.  E. B. was supported
by a NASA GSRP Graduate Fellowship under grant NGT-51662 and C. M. was
supported by an NSF Graduate Research Fellowship. G. U. thanks the
Fannie and John Hertz Foundation for fellowship support.

\end{document}